# CAN INSIDER TRADING BE COMMITTED WITHOUT TRADING?

Russell Stanley Q. Geronimo[*]

*ABSTRACT*

*Before a person can be prosecuted and convicted for insider trading, he must first execute the overt act of trading. If no sale of security is consummated, no crime is also consummated. However, through a complex and insidious combination of various financial instruments, one can capture the same amount of gains from insider trading without undertaking an actual trade. Since the crime of insider trading involves buying or selling a security, a more sophisticated insider can circumvent the language of the Securities Regulation Code by replicating the economic equivalent of a sale without consummating a sale as defined by law.*

Through the use of financial derivatives in the form of options, swaps, and forwards, an insider who is not a shareholder in a company can obtain economic exposure to changes in the market value or price of shares of stock, without purchasing or obtaining ownership of the shares. The actual stockholder or dealer of security transfers his economic exposure to the insider, but retains all stockholder rights. The insider obtains returns associated with the share of stock by assuming the financial risks inherent in stock ownership, while the person holding the shares of stock is insulated from such risks.

This paper demonstrates how constructive trades circumvent the insider trading law by allowing an insider to obtain economic exposure over a share of stock without obtaining or divesting his title over the stock.

**Introduction**

Before a person can be prosecuted and convicted for insider trading, he must first execute the overt act of trading.[1] If no sale of security is consummated, no crime is also consummated.[2] However, through a complex and insidious combination of various financial instruments, one can capture the same amount of gains from insider trading without undertaking a trade.[3] Since the crime of insider trading involves buying or selling a security, a more sophisticated insider can circumvent the language of the Securities Regulation Code by replicating the economic equivalent of a sale without consummating a sale as defined by law.[4]

---

[*] Juris Doctor, University of the Philippines College of Law, Philippines
[1] Section 27.1 of the Securities Regulation Code (R.A. No. 8799)
[2] Article 6 of the Revised Penal Code (Act No. 3815) states that "[a] felony is consummated when all the elements necessary for its execution and accomplishment are present."

Through the use of financial derivatives[5] in the form of options[6], swaps[7], and forwards[8], an insider who is not a shareholder in a company can obtain economic exposure to changes in the market value or price of shares of stock, without purchasing or obtaining ownership of the shares.[9] The actual stockholder or dealer of security transfers his economic exposure to the insider, but retains all stockholder rights.[10] The insider obtains returns associated with the share of stock by assuming the financial risks inherent in stock ownership, while the person holding the shares of stock is insulated from such risks.[11] We shall call this "constructive trade"[12], as opposed to an actual trade.

---

3   Thel, Steve, *Closing a Loophole: Insider Trading in Standardized Options*, 16 Fordham Urb. LJ 4 (1987)

4   Wang, W.K., *A Cause of Action for Option Traders Against Insider Option Traders*, 101 Harvard Law Review 5, 1056 (1988)

5   Rule 3.1.9 of the 2015 Implementing Rules and Regulation of the Securities Regulation Code states, "Derivative is a financial instrument whose value changes in response to changes in a specified interest rate, security price, commodity price, foreign exchange rate, index of prices or rates, credit rating or credit index, or similar variable or underlying factor. It is settled at a future date"; BSP Circular No. 594 series of 2008 states, "derivative is broadly defined as a financial instrument that primarily derives its value from the performance of an underlying variable."

6   *Sps. Litonjua vs. L&R Corporation* (G.R. No. 130722, March 27, 2000) defines an option as the "choice granted to another for a distinct and separate consideration as to whether or not to purchase a determinate thing at a predetermined fixed price."

7   *BDC Fin. LLC v. Barclays Bank PLC* (2012 NY Slip Op 33758)

8   *Commodity Futures Trading Com'n v. Zelener*, 387 F.3d 624 (7th Cir. 2004)

9   Knoll, M.S., *Put-call Parity and the Law*, 24 Cardozo L. Rev. 61 (2002)

10  Hu, Henry T. C. and Black, Bernard S., *The New Vote Buying: Empty Voting and Hidden (Morphable) Ownership*, 79 Southern California Law Review 811 (2006)

11  There must be no agency, partnership, trust or nominee arrangement between the insider and the stockholder; if such an agreement exists, then the stockholder is either a co-principal or accomplice of the insider, which would put the transaction between the insider and the stockholder within the purview of the insider trading law.

12  A similar concept can be found in U.S. tax regulation, called "constructive sales", involving hedging transactions and offsetting positions, often with the aid of financial derivatives. While no actual transfer of ownership takes place, certain transactions are considered sale under law to prevent abusive tax avoidance and deferral schemes. See Paul, W.M., *Constructive Sales Under New Section 1259*, 76 Tax Notes 1467 (1997).

**Economic Interest Without Ownership**

The common denominator underlying options, swaps and forwards is their ability to create economic interest without ownership.[13]

**A. Options**

Through an option contract, a stockholder or dealer agrees to give another party the right, but not the obligation, to purchase shares of stock at a future date and at a fixed price.[14] Upon the arrival of the future date, three scenarios are possible: the price stipulated in the option is higher, lower or equal to the prevailing share price.[15] If the option-holder is an insider, he has unfair advantage in the market because he possesses material non-public information that share prices will considerably increase at a future date.[16] And if indeed share prices increase after the publication of the information, the option-holder *exercises* the option.[17] However, as insider, the option-holder is criminally prohibited from executing a sales contract over the underlying shares.[18] Hence, the option-holder simply agrees to receive from the stockholder an amount representing the difference between the prevailing share price and the fixed price stipulated in the option.[19] The option-holder effectively exacts the amount of profits that he would obtain had he purchased the shares of stock directly from the stockholder or dealer before share prices increased.[20]

**B. Swaps**

Through a swap contract, one party who holds a security agrees with another party holding another security to exchange the cash flow of their respective securities.[21] Thus, if X holds a bond and Y holds a share of stock, X transfers the yield of the bond to Y, and Y transfers the return on the share of stock to X.[22] In this scenario, X (who is a bondholder) assumes the risk in the fluctuation of share prices over time, while Y (who is the stockholder) is insulated from that risk and obtains a fixed return equal to the yield of the bond.[23] Meanwhile, X does not obtain title over the shares of stock and does not divest his ownership of the bond, while Y does not obtain title over the bond and does not divest his ownership of the shares.[24] If X is an insider with respect to the issuer of the shares held by Y, X can execute a constructive trade involving a swap contract to capture gains from insider trading without trading shares of stock.[25]

---

13　Hu, Henry TC, and Bernard Black, *Hedge Funds, Insiders, and the Decoupling of Economic and Voting Ownership: Empty Voting and Hidden (Morphable) Ownership*, 13 JOURNAL OF CORPORATE FINANCE 2, 343 (2007)

14　*Gill v. Easebe Enters. (In re Easebe Enters.)*, 900 F.2d 1417 (9th Cir. 1990)

15　*Supra* note 9.

16　Section 27.1 of the Securities Regulation Code (R.A. No. 8799)

17　*Supra* note 9.

18　Section 27.2 of the Securities Regulation Code (R.A. No. 8799)

19　This is called "cash settlement". *See, e.g., Republic National Bank v. Hales*, 75 F. Supp.2d 300, S.D.N.Y. 1999)

20　This is called "synthetic equity". *See, e.g., Johnson v. Couturier*, 572 F.3d 1067 (9th Cir. 2009)

21　*See, e.g., Korea Life Insurance Co., LTD v. Morgan Guaranty Trust* (269 F. Supp.2d 424, S.D.N.Y. 2003)

22　*See, e.g., Corre Opportunities Fund, LP v. Emmis Commc'ns Corp.*, No. 14-1647 (7th Cir. Jul 02, 2015)

**C.   Forwards**

Through a forward contract, one party obligates himself to purchase or sell a security at a fixed price to be paid at a fixed future date.[26] While an option gives the holder a right, without the corresponding obligation, to buy or sell underlying shares of stock, a forward contract obligates the parties to buy or sell in the future.[27] Upon the arrival of the stipulated future date, the parties are compelled to enter into a sale.[28] However, if one of the parties is an insider with respect to the issuer of the underlying shares in the forward contract, they settle their obligations by having the seller pay the prevailing price of the shares, instead of delivering the actual shares of stock to the buyer. And since the buyer is obligated to pay the stipulated price in the forward contract, the seller will just pay the difference between the prevailing share price and the stipulated price, under the principle of legal compensation.[29] The buyer does not acquire title over the shares, but he effectively obtains economic exposure to the movements in the price of the shares.[30]

Our purpose in this article is to demonstrate how constructive trades circumvent the insider trading law by allowing an insider to obtain economic exposure over a share of stock without obtaining or divesting his title over the stock.[31] We shall demonstrate this through a specific scheme of constructive trade that uses a loan agreement coupled with two option contracts.

---

23  *Supra* note 9.
24  *Supra* note 13.
25  *Supra* note 10.
26  *See*, e.g., *Grain Land Coop v. Kar Kim Farms, Inc.,* 199 F.3d 983 (8th Cir. 1999)
27  *Abrams v. Oppenheimer Government Securities, Inc.*, 589 F. Supp. 4 (N.D. Ill. 1983)
28  *Id.*
29  This is an example of a cash-settled forward contract. *See*, e.g., *Levion v. Generale*, 822 F.Supp.2d 390 (S.D.N.Y. 2011)
30  *See*, generally, *Caiola v. Citibank, N.A.*, 295 F.3d 312 (2d Cir. 2002)
31  *See*, e.g., *Frank Lyon Company v. United States* (435 U.S. 561, 1978)

**Existing Insider Trading Regulation**

Let us say that during a board meeting at the start of the first quarter of the current year, the CEO of ABC Mining Corporation reported the discovery of a mining lode that can double the earnings of the company for the next 5 to 10 years.[32] X is a non-holder of ABC shares. Z, a member of the Board of Directors, discloses the "tip" exclusively to X, but the information is not scheduled for public announcement until next year.[33]

Insiders are either primary or secondary.[34] A primary insider has actual knowledge of material non-public information by virtue of his position of power or importance in the issuer corporation.[35] On the other hand, a secondary insider is a person who does not hold such position or who is not formally affiliated with the issuer, but who learns material non-public information from a primary insider.[36] In this illustration, Z is a primary insider and X is a secondary insider, pursuant to Section 3.8 of the Securities Regulation Code (R.A. No. 8799), which includes in the definition of an insider "a director […] or […] a person who learns such information by a communication from any of the foregoing insiders."

The "tip" given by Z to X is material non-public information, pursuant to Section 27.2 of the Securities Regulation Code (R.A. No. 8799), which states that "information is 'material nonpublic' if: (a) It has not been generally disclosed to the public and would likely affect the market price of the security after being disseminated to the public and the lapse of a reasonable time for the market to absorb the information; or (b) would be considered by a reasonable person important under the circumstances in determining his course of action whether to buy, sell or hold a security."

The information is not "generally available" to the public because it has not been disclosed in any form of media except through a report by senior management to the Board of Directors during a board meeting. The Supreme Court in *SEC vs. Interport Resources Corp.*[37] states that information is "generally available" to the public if "found in a newspaper, a specialized magazine, or any cyberspace media" or "made known to the public through any form of media."

---

32  For the relationship between the business model of a mining company and prices of mining stocks, see Poskitt, R., *Are Mining-Exploration Stocks More Prone to Informed Trading Than Mining-production Stocks?*, 30 AUSTRALIAN JOURNAL OF MANAGEMENT 2, 201 (2005)

33  Presentation of earning prospects by the corporation to the public is usually conducted through annual earnings announcements. *See* Beaver, W.H., *The Information Content of Annual Earnings Announcements*, JOURNAL OF ACCOUNTING RESEARCH 67 (1968).

34  Section 3.8 of the Securities Regulation Code (R.A. No. 8799)

35  *Id.*

36  *Id.*

37  G.R. No. 135808, October 06, 2008

The information is "material" because a public announcement is likely to increase share prices, and that a reasonable person would consider the projected income stream of the corporation as a factor in deciding whether to buy or sell ABC shares. Again, the Court in *SEC vs. Interport Resources Corp.* states, "A discussion of the 'materiality concept' would be relevant to both a material fact which would affect the market price of a security to a significant extent and/or a fact which a reasonable person would consider in determining his or her cause of action with regard to the shares of stock."

Possession of material non-public information prohibits X from purchasing, and Z from selling, ABC shares before public disclosure of the material information, pursuant to Section 27.1 of the Securities Regulation Code (R.A. No. 8799), which states:

It shall be unlawful for an insider to sell or buy a security of the issuer, while in possession of material information with respect to the issuer or the security that is not generally available to the public[.]

There are two modes of committing insider trading. First, where the insider is already an existing holder of a share of stock or security, the insider averts losses by selling the security, expecting price to drop after publication or announcement of the material information.[38] Second, where the insider is not a holder of the share or security, the insider makes a gain by buying the share or security, expecting price to increase after publication or announcement.[39] These prohibited overt acts in Section 27.1 are illustrated as follows:

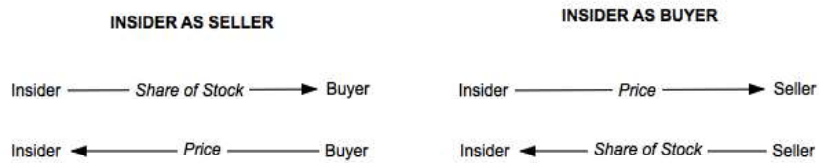

It is not necessary to realize the gains in order to consummate insider trading. It is enough that the insider buys or sells the security of the issuer while in possession of material non-public information.[40]

---

38 Section 27.1 of the Securities Regulation Code (R.A. No. 8799)
39 Section 27.1 of the Securities Regulation Code (R.A. No. 8799)
40 Profit is not an element of insider trading under Section 27.1, R.A. No. 8799.

The insider who is a seller transfers ownership of shares of stock and delivers certificates of stock to the buyer.[41] The buyer pays the selling price based on market value and the insider averts the imminent loss on the price of the shares.[42] On the other hand, the insider who is a buyer obtains ownership and possession of the shares of stock, and pays the selling price, also based on current market value.[43] The insider obtains unrealized gains from the price increase, and he needs to sell, transfer, exchange or dispose the purchased shares in order to realize those gains.[44]

**Insider Trading Without Trading**

Now assume that the indicative price increase of ABC shares after public announcement is from ₱100 to ₱150 per share, for a gain of 50%.[45] If there were no insider trading law, X would exploit the information by buying 1 million ABC shares at the current price of ₱100 per share, and by selling the same number of shares at the increased price of ₱150 per share after public disclosure of material information.[46] The hypothetical transaction is summarized as follows:

|                                              | Per Share (₱) | Total (₱)    |
| -------------------------------------------- | ------------- | ------------ |
| Selling Price (after announcement)           | 150           | 150,000,000  |
| Less: Acquisition Cost (before announcement) | 100           | 100,000,000  |
| Gain from Insider Trading                    | 50            | 50,000,000   |

As insider, X does not want to incur criminal liability for trading ABC shares.[47] Nevertheless, he wants to profit from the expected increase in share price without executing an actual trade.[48] Suppose that, being a sophisticated investor, X knows how to combine financial instruments in such a way that he can capture or mimic the gains in a share of stock without trading that stock.[49]

---

41   Observe, however, the concept of "scripless" securities in the trading of publicly listed securities (*San Miguel Corp. vs. Corporation Finance Department*, SEC *En Banc* Case No. 10-10-219, 09 December 2010)

42   The insider is "bearish". (*U.S. v. Heron*, 525 F. Supp.2d 729, E.D. Pa. 2007)

43   The insider is "bullish". (*State v. Plummer*, 117 N.H. 320, N.H. 1977)

44   For realization of unrealized gains, *see*, e.g., *S.R.G. Corp. v. Department of Revenue*, 365 So.2d 687 (Fla. 1978)

45   For the effect of announcement of discovery of mining lode in price of mining stocks, *see*, e.g., *SEC v. Texas Gulf Sulphur*, 312 F. Supp. 77 (S.D.N.Y. 1970)

46   *Supra* note 43.

47   *Supra* note 1.

48   Arnold, T., Erwin, G.R., Nail, L.A. and Bos, T., *Speculation or Insider Trading: Informed Trading in Options Markets Preceding Tender Offer Announcements* (2000), available at SSRN 234797.

49   Lin, J.C. and Howe, J.S., *Insider Trading in the OTC Market*, 45 THE JOURNAL OF FINANCE 4, 1273 (1990)

Assume that the current date, which is before the announcement, is 04 January 2015, and that the date of the announcement is 04 January 2016. Next, we will add a third party named Y, which is a hedge fund, broker, financial institution or dealer in securities.[50] The constructive trade is executed between X and Y through the following contracts, with their respective terms and stipulations:

1. A loan agreement[51] extended by X to Y, with a principal amount equal to what the original acquisition would be if X purchased the ABC shares;

2. An option to buy[52] the ABC shares, giving X the right, but not the obligation, to purchase ABC shares, with strike price equal to the full amount of the loan at maturity (i.e. principal and interest) and with exercise date on or after the date of publication or announcement of the material information; and

3. An option to sell[53] ABC shares, giving Y the right, but not the obligation, to sell ABC shares, with strike price equal to the full amount of the loan at maturity (i.e. principal and interest) and with exercise date on or after the date of publication or announcement of the material information.

We shall discuss each of these instruments in the following sections.

**The Loan Agreement**

In an actual trade, the insider pays the purchase price to obtain ABC shares from an existing stockholder, who transfers the shares to the insider.[54] In a constructive trade, the insider grants a loan, with a principal amount equal to what the amount of the purchase price would be if the insider executes an actual trade.[55] Through constructive trade, X (the insider) becomes a creditor of Y (the dealer in security who is not an insider). The initial transaction is illustrated as follows:

---

50 For the participation of hedge funds and dealers in securities in the transfer of risk-return profile of securities without actual transfer of ownership, *see* Hu, H.T. and Black, B., *Hedge Funds, Insiders, and the Decoupling of Economic and Voting Ownership: Empty Voting and Hidden (Morphable) Ownership*, 13 JOURNAL OF CORPORATE FINANCE 2, 343 (2007)

51 Schlunk, H.J., *Little Boxes: Can Optimal Commodity Tax Methodology Save the Debt-Equity Distinction*, 80 TEX. L. REV. 859 (2001)

52 Knoll, M.S., *Financial Innovation, Tax Arbitrage, and Retrospective Taxation: The Problem with Passive Government Lending*, 52 TAX L. REV. 199 (1996)

53 Gergen, M.P., *Afterword Apocalypse Not*, 50 TAX L. REV. 833 (1994)

54 Such would be covered by a Stock Purchase Agreement (*Romago Electric Co. vs. CA*, G.R. No. 125947, June 08, 2000)

55 The loan represents any debt instrument. In finance, the traditional instrument is a zero-coupon bond. A loan with zero interest replicates the economic characteristic of a zero-coupon bond. (Bossu, S., *Put–Call Parity*, ENCYCLOPEDIA OF QUANTITATIVE FINANCE (2010))

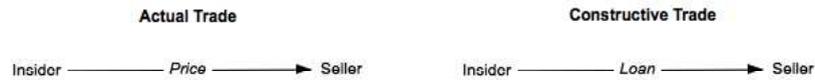

On 04 January 2015, the current share price is ₱100 per share, so that a block of 1 million ABC shares is equivalent to ₱100 million. Instead of buying 1 million ABC shares (which would fall within the purview of insider trading), X lends a sum of money to Y. The sum of money represents the principal amount of ₱100 million, which is equal to the current purchase price of 1 million ABC shares at ₱100 per share. X and Y also stipulate that the interest on the loan is 5%.

**The Option to Buy**

After the loan, X and Y execute an option contract over the ABC shares. The Supreme Court in *Eulogio vs. Sps. Apeles*[56] defines an option as "a contract by which the owner of the property agrees with another person that the latter shall have the right to buy the former's property at a fixed price within a certain time."

**More formally, an option contract has the following elements:**

(1)   Underlying asset[57]

(2)   The right, but not the obligation, to buy the underlying asset[58]

(3)   Strike price[59]

(4)   Exercise date[60]

(5)   Option style[61]

The underlying asset consists of 1 million ABC shares. The option has two parties: the owner[62] of property and the holder of the option. The option-holder is the party given the right, but not the obligation, to buy the property.[63] In this case, Y is the owner and X is the option-holder. Since the option involves a right to buy, it is a call option.[64]

---

56   G.R. No. 167884, January 20, 2009
57   *Ruppert v. Alliant Energy Cash Balance Pension Plan*, 08-cv-127-bbc. (W.D. Wis. Dec 29, 2010)
58   *Deutschman v. Beneficial Corp.*, 668 F. Supp. 358 (D. Del. 1987)
59   Ross, S.A., *Options and Efficiency*, The Quarterly Journal of Economics 75 (1976)
60   *Steele v. Northup*, 259 Iowa 443 (Iowa 1966)
61   *Hollingshad v. Deutsche Bank AG*, Civil Action No. 3:05-CV-2235-L., N.D. Tex. Oct 03, 2006
62   In actual commercial practice, a party to an option giving the option-holder a right to buy the underlying asset does not even have to be the owner at the time of execution of the option contract. *See*, e.g., *People v. Daman*, C069199 (Cal. Ct. App. Feb 25, 2013).
63   *Deutschman v. Beneficial Corp.*, 841 F.2d 502 (3d Cir. 1988)
64   *Moskowitz v. Lopp*, 128 F.R.D. 624 (E.D. Pa. 1989)

The strike price is the fixed future price at which the option-holder would buy the underlying asset should he decide to exercise the option.[65] The exercise date is the date when the option-holder can decide to avail of his right to buy the underlying asset.[66] If the exercise date lapses and the option-holder failed to exercise the option, the option contract expires.[67]

The option style refers to the period during which the option-holder can exercise the option.[68] The style can be European or American. Under the European style, the holder can only exercise the option at (and not before) the agreed exercise date. Under the American style, the holder can exercise the option any time before and at the agreed exercise date.[69]

For the purpose of our illustration, let the strike price be ₱105 per share, or a total of ₱105 million. Let the exercise date be at 04 January 2016, which is the date of the announcement. Finally, let the call option between X and Y be a European-style option, i.e. capable of being exercised only upon the arrival of 04 January 2016. To summarize, the parameters of the call option are as follows:

| Option-Holder | X |
| --- | --- |
| Owner of Underlying Asset | Y |
| Underlying Asset | 1 million ABC shares |
| Style | European |
| Strike Price | ₱105 per share, for a total of ₱105 million |
| Exercise Date | 04 January 2016 |

It is essential that the strike price of the call option must be equivalent to the full amount of the loan at maturity, i.e. the principal amount *plus* the amount of interest.[70]

Before we move to the next step, we must clarify the legal nature of the call option contract between X and Y. Since the call option gave X the right to buy the ABC shares, the question is whether X had engaged in "trading". Is the execution of a call option contract, *even before the exercise of the option by X*, considered a "purchase" of the underlying asset, for which X would be liable for insider trading?

---

65  *Wasserman v. Triad Securities Corp.*, Case No. 8:05-cv-1898-T-24TBM. (M.D. Fla. Jun 12, 2006)
66  *Allaire Corp. v. Okumus*, 433 F.3d 248 (2d Cir. 2006)
67  *Lerner v. Millenco, L.P.*, 23 F. Supp.2d 337 (S.D.N.Y. 1998)
68  For examples of different option styles, *see*, e.g., *New Millennium Trading, LLC v. Commisioner*, 131 T.C. 275 (T.C. 2008)
69  There are other styles aside from American and European. *See*, e.g., *Thomas Investment Partners, LTD v. U.S.* (Nos. 09-55638, 09-55639, 09-55641, 09-55642, 09-55650., 9th Cir. Jul 20, 2011)
70  *Supra* note 9.

**Execution of Option Contract**

The overt act of insider trading is explicitly defined in Section 27.1 of the Securities Regulation Code (R.A. No. 8799), which states that "[i]t shall be unlawful for an insider to sell or buy a security of the issuer[.]" The Supreme Court in *Eulogio vs. Sps. Apeles*[71], however, states that "[a]n option is not of itself a purchase, but merely secures the privilege to buy. It is not a sale of property but a sale of the right to purchase." Moreover, the Supreme Court in *Sps. Litonjua vs. L&R Corp.*[72] states, "Observe, however, that the option is not the contract of sale itself."

A contract of sale has three stages: negotiation, perfection and consummation. The Supreme Court in *Manila Metal Container Co. vs. PNB*[73] states:

[T]he stages of a contract of sale are as follows: (1) *negotiation*, covering the period from the time the prospective contracting parties indicate interest in the contract to the time the contract is perfected; (2) *perfection*, which takes place upon the concurrence of the essential elements of the sale which are the meeting of the minds of the parties as to the object of the contract and upon the price; and (3) *consummation*, which begins when the parties perform their respective undertakings under the contract of sale, culminating in the extinguishment thereof.

*Prior to the exercise*[74] *of the option*, the sale of the shares of stock was not yet perfected. In order for a sale to be perfected, there must be a meeting between offer and acceptance, pursuant to Article 1319 of the New Civil Code, which states that "[c]onsent is manifested by the meeting of the offer and the acceptance upon the thing and the cause which are to constitute the contract. The offer must be certain and the acceptance absolute. A qualified acceptance constitutes a counter-offer."

In the mere execution[75] of an option contract (i.e. prior to the exercise of the option by the holder), there is still no concurrence of offer and acceptance. The

---

71 G.R. No. 167884, January 20, 2009
72 G.R. No. 130722, March 27, 2000
73 G.R. No. 166862, December 20, 2006
74 Not to be confused with "execute". Execution entails the creation of the option contract. To exercise the option means that the option-holder decides to avail of his right to buy or sell the underlying asset of the option, within the period or at the time stipulated in the option contract.
75 Not to be confused with "exercise".

Supreme Court in *Adelfa Properties vs. Court of Appeals*[76] states that "[a]n option, as used in the law on sales, is a continuing offer […] It is also sometimes called an 'unaccepted offer.'"

**The Supreme Court in *Asuncion vs. CA*[77] states:**

The option, however, is an independent contract by itself, and it is to be distinguished from the projected main agreement (subject matter of the option) which is obviously yet to be concluded. If, in fact, the optioner-offeror withdraws the offer before its acceptance (exercise of the option) by the optionee-offeree, the latter may not sue for specific performance on the proposed contract ("object" of the option) since it has failed to reach its own stage of perfection.

Article 1479 of the New Civil Code defines an option contract as a unilateral promise supported by a consideration distinct from the price. It states:

An accepted unilateral promise to buy or to sell a determinate thing for a price certain is binding upon the promissor if the promise is supported by a consideration distinct from the price.

The Supreme Court in *Eulogio vs. Sps. Apeles*[78] distinguishes between the sale of an asset, on the one hand, and the sale of the right to buy the underlying asset, on the other. The Court provides:

He does not sell his land; he does not then agree to sell it; but he does sell something, *i.e.,* the right or privilege to buy at the election or option of the other party. Its distinguishing characteristic is that it imposes no binding obligation on the person holding the option, aside from the consideration for the offer.

Is the sale of the right or privilege to buy a stock considered trading the stock? A strict construction of Section 27.1 of the Securities Regulation Code shows that this is not covered by the prohibition on insider trading. The provision states that "[i]t shall be unlawful for an insider to sell or buy a security of the issuer[.]" The purchase of a security is not equivalent to the purchase of a right to purchase the security. Although Section 3.1 of the Securities Regulation Code states that "securities" include options, Section 27.1 of the the same Code states that insider trading involves the selling or buying of the security "of the issuer", and the option contract in this case is not issued by ABC Inc. It is a private contract between X and Y. Accordingly, execution of an option contract does not constitute "trading" under Section 27.1.

---

76   G.R. No. 111238, January 25, 1995
77   G.R. No. 109125, December 02, 1994
78   G.R. No. 167884, January 20, 2009

So far, X and Y executed two agreements: a loan contract and a call option contract. The next step is the execution of another option, whereby X gives Y the right, but not the obligation, to sell 1 million ABC shares to X.

**The Option to Sell**

The Supreme Court in *Limson vs. Court of Appeals*[79] states that an option contract may also give to the owner of the property "the right to sell or demand a sale". This type of option contract is called a "put option", and is the opposite of a "call option".[80]

Under a call option, the option-holder has the right, but not the obligation, to buy the property. Under a put option, the owner (who is the option-holder) has the right, but not the obligation, to sell the property to a non-owner.[81]

Under a call option, the owner has the unconditional obligation to sell the property the moment the option-holder exercises his option to buy. Under a put option, the non-owner has the unconditional obligation to buy the property the moment the owner (who is the option-holder) exercises his option to sell.[82]

The put option gives Y the right, but not the obligation, to sell 1 million ABC shares to X on 04 January 2016 at a strike price of ₱105 per share, or a total of ₱150 million. It has the following parameters:

| Option-Holder | Y |
| --- | --- |
| Counterparty | X |
| Underlying asset | 1 million ABC shares |
| Option style | European |
| Strike price (₱) | ₱105 per share, for a total of ₱105 million |
| Exercise date | 04 January 2016 |

The strike price of the put option must be equivalent to the strike price of the call option, which in turn is equivalent to the principal amount of the loan *plus* interest.[83]

---

[79] G.R. No. 135929, April 20, 2001
[80] *Gentile v. Rossette*, C.A. No. 20213 (Del. Ch. May 28, 2010)
[81] *Madison International Liquidity Fund, LLC v. Al. Neyer*, Case No. C-1-08-305. (S.D. Ohio Jul 17, 2009)
[82] *Anserphone of New Orleans, Inc. v. Protocol Comm.*, Civil Action No. 01-3740, Section "A" (1) (E.D. La. Dec 05, 2002)
[83] *Supra* note 9.

The following is a summary of the equivalent values:

$$\text{Beginning Share Price} = \text{Principal Amount of Loan}$$

$$\text{Strike Price of Call Option} = \text{Strike Price of Put Option}$$
$$= \text{Principal Amount of Loan} + \text{Interest}$$

In the example, the share price as of 04 January 2015 (the beginning date) is ₱100 million, which is equal to the principal amount of the loan. And the strike price of each option is equal to ₱105 million, which is equal to the loan principal amount of ₱100 million and the interest of ₱5 million upon maturity.

Aside from the equality of strike prices, the call and put option must have the same underlying asset (1 million ABC shares), the same exercise date (04 January 2016), and the same style (European). Moreover, the exercise date of either option must also be the same as the date of maturity of the loan. These relationships of equality are essential to the constructive trading scheme of X, to be discussed in the following sections.

**Exercising the Option**

Now assume that on 04 January 2016, upon public disclosure of the material information, there is an instantaneous reaction in the financial markets and the price of ABC shares accordingly adjusts to the news.[84] Three scenarios can occur: (1) the new share price is higher than the strike price of either option, (2) lower than the strike price, or (3) equivalent to the strike price.

If the unrealized gain is ₱50 per share, the new share price is ₱150 per share, computed as follows:

|  | Per Share (₱) | Total (₱) |
| --- | --- | --- |
| Share Price (04 January 2015) | 100 | 100,000,000.00 |
| Less: Unrealized Gain | 50 | 50,000,000.00 |
| Share Price (04 January 2016) | 150 | 150,000,000.00 |

Since the new share price of ₱150 per share is higher than the strike price of ₱105 per share, X exercises the call option, for a profit of ₱45 per share in favor of X equivalent to the difference between the new share price and the strike price. X has an obligation to pay the strike price and Y has the obligation to deliver the shares. This is illustrated as follows:

---

84  For the relationship between stock prices and news, *see* McQueen, G. and Roley, V.V., *Stock Prices, News, and Business Conditions*, 6 REVIEW OF FINANCIAL STUDIES 3, 683 (1993)

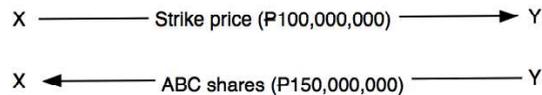

What is the legal implication of exercising the option contract? Will it not constitute "trading" for purposes of the insider trading law?

**Legal Implication of Exercising the Option**

Under the law on sales, the perfection of a sale is different from its consummation.[85] Exercising the option constitutes the perfection of the sale, but delivery of the object of sale constitutes the consummation. Therefore, strictly speaking, the act of exercising the option can be made without a subsequent delivery of the shares from Y (seller) to X (buyer), and it is possible to perfect the sale without consummating it later on.

We shall elaborate this by recalling the three stages of a contract. The Supreme Court in *C.F. Sharp & Co. Inc. vs. Pioneer Insurance & Surety Corp.*[86] states:

[C]ontracts undergo three distinct stages, to wit: negotiation; perfection or birth; and consummation. Negotiation begins from the time the prospective contracting parties manifest their interest in the contract and ends at the moment of agreement of the parties. Perfection or birth of the contract takes place when the parties agree upon the essential elements of the contract. Consummation occurs when the parties fulfill or perform the terms agreed upon in the contract, culminating in the extinguishment thereof.

**This is illustrated as follows:**

|  | Stages of Sale | | |
| --- | --- | --- | --- |
|  | Negotiation | Perfection | Consummation |
| Acts of the Parties | X and Y execute the option contract | X exercises the option contract | X pays the strike price of ₱105 million, and Y delivers the shares of stock valued at ₱150 million |
| Juridical Nature of the Acts | Y extends a unilateral promise to sell[87]; an unaccepted offer[88] | X and Y have a bilateral promise to buy and sell; meeting of offer and acceptance[89] | X and Y perform their respective obligations[90] |

---

85  *Sps. Serrano and Herrera vs. Caguiat*, G.R. No. 139173, February 28, 2007
86  G.R. No. 179469, February 15, 2012
87  Article 1479 of the New Civil Code
88  *Reyes vs. CA*, G.R. No. 94214, December 01, 1992
89  *Sps. Litonjua vs. L&R Corp.*, G.R. No. 130722, March 27, 2000
90  *Far East Bank and Trust Company vs. PDIC*, G.R. No. 172983, July 22, 2015

The execution of the option contract constitutes the negotiation stage, pursuant to Article 1479 of the New Civil Code, which defines the option contract as a unilateral promise to sell. The Supreme Court in *Atkins, Kroll & Co., Inc. vs. Cua Hian Tek*[91] also states that "[a]fter accepting the promise and before he exercises his option, the holder of the option is not bound to buy. He is free either to buy or not to buy later."

Exercising the option contract constitutes the perfection stage. At this point, X has accepted the unilateral promise of Y. The meeting of offer and acceptance yields to the bilateral promise to buy and sell. The Supreme Court in *Sps. Litonjua vs. L&R Corp.*[92] states that "[o]nce the option is exercised timely, i.e., the offer is accepted before a breach of the option, a bilateral promise to sell and to buy ensues and both parties are then reciprocally bound to comply with their respective undertakings."

In other words, at the perfection stage, there already exists a contract of sale. The Supreme Court in *Atkins, Kroll & Co., Inc. vs. Cua Hian Tek*[93] states that "upon accepting herein petitioner's offer[,] a bilateral promise to sell and to buy ensued, and the respondent *ipso facto* assumed the obligations of a purchaser." This is supported by the Concurring Opinion in *Sanchez vs. Rigos*[94], stating that "[i]f […] acceptance is made before a withdrawal, it constitutes a binding contract of sale. The concurrence of both acts – the offer and the acceptance – could in such event generate a contract."

Finally, the performance of the respective obligations of X and Y constitutes the consummation stage of the sale. The obligation of X is to pay the strike price, and the obligation of Y is to deliver the shares of stock. The Supreme Court in *Far East Bank and Trust Company vs. PDIC*[95] states that "[t]he consummation stage begins when the parties perform their respective undertakings under the contract, culminating in its extinguishment."

And what are their respective undertakings? Article 1458 of the New Civil Code states that "[b]y the contract of sale one of the contracting parties obligates himself to transfer the ownership and to deliver a determinate thing, and the other to pay therefor a price certain in money or its equivalent." The Supreme Court in *ACE Goods, Inc. vs. Micro Pacific Technologies*[96] states that "[t]he very essence of a contract of sale is the transfer of ownership in exchange for a price paid or promised."

---

91  G. R. No. L-9871, January 31, 1958
92  G.R. No. 130722, March 27, 2000
93  G.R. No. L-9871, January 31, 1958
94  G.R. No. L-25494, June 14, 1972
95  G.R. No. 172983, July 22, 2015
96  G.R. No. 200602, December 11, 2013

With respect to the transfer of ownership of shares of stock, the Supreme Court in *Fil-Estate Golf and Development, Inc. vs. Vertex Sales and Trading, Inc.*[97] states that "[in a] sale of shares of stock, physical delivery of a stock certificate is one of the essential requisites for the transfer of ownership of the stocks purchased." Section 63 of the Corporation Code provides:

Shares of stock so issued are personal property and may be transferred by delivery of the certificate or certificates indorsed by the owner or his attorney-in-fact or other person legally authorized to make the transfer. No transfer, however, shall be valid, except as between the parties, until the transfer is recorded in the books of the corporation showing the names of the parties to the transaction, the date of the transfer, the number of the certificate or certificates and the number of shares transferred.

If X exercises the call option, the act perfects the contract of sale over the shares of stock, and X and Y are now bound by a bilateral promise to buy and sell.[98] However, X and Y are prohibited from consummating their respective obligations—i.e. X is prohibited from paying the strike price to Y, while Y is prohibited from delivering the shares of stock to X.[99] If X acquires title over the shares, he consummates the proscribed activity of insider trading.[100]

**How to Settle the Option Without Transfer of Shares**

There are two ways to settle an option contract: physical delivery of shares or cash netting arrangement.[101] Physical delivery of shares requires the transfer of the shares of stock in the name of the buyer.[102] Cash netting arrangement, on the other hand, is a mode of settlement whereby the seller, who promised to deliver the shares to the buyer, will instead pay the fair market value of the said shares.[103] Since the fair market value is paid through a sum of money, the parties offset the strike price with the payment representing the shares' fair market value, so that the seller will only pay the difference between the fair market value and the strike price if the fair market value is higher.[104] On the other hand, the buyer will pay the difference to the seller if the strike price is higher than the fair market value.[105]

---

97  G.R. No. 202079, June 10, 2013
98  *Sps. Litonjua vs. L&R Corp.,* G.R. No. 130722, March 27, 2000
99  Section 27.1, R.A. No. 8799
100 Section 2, R.A. No. 8799
101 Feder, N.M., *Deconstructing Over-the-Counter Derivatives*, COLUM. BUS. L. REV. 677 (2002)
102 *Morris v. Kaiser*, 292 Ala. 650 (Ala. 1974)
103 *Republic National Bank v. Hales*, 75 F. Supp.2d 300 (S.D.N.Y. 1999)
104 *Alaska Elec. Pension Fund v. Bank of Am. Corp.*, 14-CV-7126 (JMF) (S.D.N.Y. Mar 28, 2016)
105 *Id.*

In order to realize the profit of ₱45 million without transferring ownership over the shares in the name of X, the parties settle through a cash netting arrangement.[106] Since physical delivery of shares effects a transfer of ownership over the shares of stock,[107] X and Y can only settle the option through a cash netting arrangement, which entails an objective novation coupled with legal compensation.[108]

Upon the exercise of the call option by X, both parties now have the following set of rights and obligations under the loan contract and call option contract:

| Parties | Loan | Call Option upon exercise by X | |
|---|---|---|---|
| X | Right to receive principal amount of ₱100 million, and interest of ₱5 million, for a total of ₱105 million | Obligation to pay ABC shares at strike price of ₱105 million | Right to receive ABC shares at new share price of ₱150 million |
| Y | Obligation to repay principal amount of ₱100 million, and to pay interest of ₱5 million, for a total of ₱105 million | Right to receive strike price of ₱105 million | Obligation to deliver ABC shares at new share price of ₱150 million |

Instead of transferring title over ABC shares,[109] which would consummate a "trade", X and Y agree to settle the call option by delivering the monetary equivalent of the ABC shares. This is in the nature of a novation.[100] Hence, we modify the table above, as follows:

| Parties | Loan | Call Option upon exercise by X | |
|---|---|---|---|
| X | Right to receive principal amount of ₱100 million, and interest of ₱5 million, for a total of ₱105 million | Obligation to pay ABC shares at strike price of ₱105 million | Right to receive the monetary equivalent of ABC shares at new share price of ₱150 million |
| Y | Obligation to repay principal amount of ₱100 million, and to pay interest of ₱5 million, for a total of ₱105 million | Right to receive strike price of ₱105 million | Obligation to deliver the monetary equivalent of ABC shares at new share price of ₱150 million |

---

106 Corbi, A., *Netting and Offsetting: Reporting derivatives under US GAAP and under IFRS*, International Swaps and Derivatives Association (ISDA), New York, USA (2012)
107 Section 63 of the Corporation Code
108 Article 1278 of the New Civil Code
109 Section 63 of the Corporation Code
110 Article 1291 of the New Civil Code

This is sanctioned by Article 1291 of the New Civil Code, which states that "[o]bligations may be modified by […] [c]hanging their object or principal conditions." The object of the call option is the right to purchase 1 million ABC shares. This was modified, under the concept of objective novation, by extinguishing "the right to purchase ABC shares" and introducing a new one, "the right to receive the monetary equivalent of ABC shares"; and by extinguishing "the obligation to deliver ABC shares" and introducing a new one, "the obligation to deliver the monetary equivalent of ABC shares".

**The Supreme Court in *Ajax Marketing & Dev't Corp. vs. CA*[111] states:**

Novation is the extinguishment of an obligation by the substitution or change of the obligation by a subsequent one which extinguishes or modifies the first […] Novation, unlike other modes of extinction of obligations, is a juridical act with a dual function, namely, it extinguishes an obligation and creates a new one in lieu of the old. It can be objective, subjective, or mixed. Objective novation occurs when there is a change of the object or principal conditions of an existing obligation[.]

With the novation, X and Y have become mutual debtors and creditors under the call option, pursuant to Article 1278 of the New Civil Code, which states that "[c]ompensation shall take place when two persons, in their own right, are creditors and debtors of each other." Article 1279 of the New Civil Code provides the requisites for legal compensation:

**In order that compensation may be proper, it is necessary:**

1. That each one of the obligors be bound principally, and that he be at the same time a principal creditor of the other;

2. That both debts consist in a sum of money, or if the things due are consumable, they be of the same kind, and also of the same quality if the latter has been stated;

3. That the two debts be due;

4. That they be liquidated and demandable;

5. That over neither of them there be any retention or controversy, commenced by third persons and communicated in due time to the debtor.

Legal compensation, as a mode of extinguishing an obligation, takes place by operation of law.[112] This modifies the table of rights and obligations between X and Y, as follows:

---

111 G.R. No. 118585, September 14, 1995
112 *UBP vs. DBP*, G.R. No. 191555, January 20, 2014

| Parties | Loan | Call Option upon exercise by X after setoff |
|---------|------|---------------------------------------------|
| X | Right to receive principal amount of ₱100 million, and interest of ₱5 million, for a total of ₱105 million | Right to receive ₱45 million[113] |
| Y | Obligation to repay principal amount of ₱100 million, and to pay interest of ₱5 million, for a total of ₱105 million | Obligation to pay ₱45 million[114] |

Based on the table, Y has a total obligation of ₱150 million, representing his obligation under the loan contract *plus* his obligation under the call option. Hence, Y pays the amount to extinguish both obligations. In doing so, he may or may not dispose the ABC shares in the market. If he disposes the ABC shares, he obtains ₱150 million as proceeds of sale in 04 January 2016, which is exactly what he needs to extinguish his total obligation to X. But if he does not dispose the ABC shares, he retains title to the ABC shares at the new market value of ₱150 million, and can pay his obligation at the same amount to X using his remaining cash. Whether or not Y disposes the ABC shares is a matter of indifference to X.

X has a positive payoff from the call option at ₱45 per share (the difference between the monetary equivalent of the new share price and the strike price). In addition, we must add to this positive payoff the amount he receives as full payment for the loan extended to Y. His economic position is as follows:

---

[113] The difference between the right of X to receive ₱150 million and an obligation to pay the strike price of ₱105 million

[114] The difference between the obligation of Y to pay ₱150 million and a right to receive the strike price of ₱105 million.

| | Per Share (₱) | Total (₱) |
|---|---|---|
| Monetary Equivalent of ABC Shares (04 January 2016) | 150 | 150,000,000.00 |
| Less: Strike Price | 105 | 105,000,000.00 |
| Payoff from Call Option | 45 | 45,000,000.00 |
| Add: Loan Receivable | | 105,000,000.00 |
| Total Amount Entitled (04 January 2016) | | 150,000,000.00 |

X's entitlement to ₱150 million represents the following: (1) repayment of the principal amount of the loan extended to Y, which is ₱100 million; (2) interest income on the loan, which is ₱5 million; and (3) gain resulting from economic exposure to the price of ABC shares, which is ₱45 million. Note that ₱150 million is exactly the total value of ABC shares in his portfolio if he purchased it from the market, in violation of the insider trading law.

Y, on the other hand, has zero net position (i.e. he neither gains nor loses any amount), computed as follows:

| Strike Price | 105 | 105,000,000.00 |
|---|---|---|
| Less: Monetary Equivalent of FMV of ABC Shares (04 January 2016) | 150 | 150,000,000.00 |
| Negative Payoff from Call Option | (45) | (45,000,000.00) |
| Add: New Share Price (04 January 2016) | 150 | 150,000,000.00 |
| Gross Position | 105 | 105,000,000.00 |
| Less: Loan Repayment | | 105,000,000.00 |
| Net Position | | 0.00 |

This illustrates the transfer of economic exposure to the price of ABC shares from Y to X. Ownership of ABC shares was not transferred to X, but he benefited from the increase in the stock price of ABC shares. And even though Y retained the stock ownership, he did not benefit from the increase in the stock price of ABC shares.

Under an ordinary buy-and-sell transaction, which is criminally prohibited in this context if executed by X, the latter would have made the same gain, as follows:

|  | Per Share (₱) | Total (₱) |
| --- | --- | --- |
| Selling Price (after announcement) | 150 | 150,000,000 |
| Less: Acquisition Cost (before announcement) | 100 | 100,000,000 |
| Gain from Insider Trading | 50 | 50,000,000 |

Through the combination of loan agreement, option to buy and option to sell the shares of stock, X's financial position in the end mimics, simulates or replicates gains from trading without engaging in an actual trade. There was no consummated sale, exchange, transfer or disposition of ABC shares in any manner.

Suppose, however, that instead of an increase in share price from 04 January 2015 to 04 January 2016, ABC shares dropped in market price. If the unrealized loss is ₱20 per share, the new share price is ₱80 per share, computed as follows:

|  | Per Share (₱) | Total (₱) |
| --- | --- | --- |
| Share Price (04 January 2015) | 100 | 100,000,000.00 |
| Less: Unrealized Loss | 20 | 20,000,000.00 |
| Share Price (04 January 2016) | 80 | 80,000,000.00 |

Since the new share price of ₱80 per share is lower than the strike price of ₱105 per share, Y exercises the put option in order to protect himself from the decrease in the share price. X has an obligation to pay the strike price and Y has the obligation to deliver the monetary equivalent of the new share price. This is illustrated as follows:

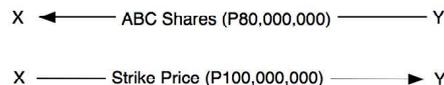

Recall that, in 04 January 2016, both parties have the following set of rights and obligations under the loan contract and put option contract:

---

96 Henry Campbell Black, BLACK'S LAW DICTIONARY (4), 519
97 Henry Campbell Black, BLACK'S LAW DICTIONARY (4), 1157
98 Henry Campbell Black, BLACK'S LAW DICTIONARY (4), 458

| Parties | Loan | Put Option upon exercise by Y | |
|---|---|---|---|
| X | Right to receive principal amount of ₱100 million, and interest of ₱5 million, for a total of ₱105 million | Obligation to pay ABC shares at strike price of ₱105 million | Right to receive ABC shares at new share price of ₱80 million |
| Y | Obligation to repay principal amount of ₱100 million, and to pay interest of ₱5 million, for a total of ₱105 million | Right to receive strike price of ₱105 million | Obligation to deliver ABC shares at new share price of ₱80 million |

Instead of transferring the title over ABC shares, X and Y agree to settle the put option by delivering the monetary equivalent of the ABC shares, as follows:

| Parties | Loan | Put Option upon exercise by X | |
|---|---|---|---|
| X | Right to receive principal amount of ₱100 million, and interest of ₱5 million, for a total of ₱105 million | Obligation to pay ABC shares at strike price of ₱105 million | Right to receive the monetary equivalent of ABC shares at new share price of ₱80 million |
| Y | Obligation to repay principal amount of ₱100 million, and to pay interest of ₱5 million, for a total of ₱105 million | Right to receive strike price of ₱105 million | Obligation to deliver the monetary equivalent of ABC shares at new share price of ₱80 million |

With the novation, X and Y have become mutual debtors and creditors under the put option. The Civil Code provision on legal compensation takes place by operation of law, as follows:

| Parties | Loan | Put Option upon exercise by X after setoff |
|---|---|---|
| X | Right to receive principal amount of ₱100 million, and interest of ₱5 million, for a total of ₱105 million | Obligation to pay ₱25 million |
| Y | Obligation to repay principal amount of ₱100 million, and to pay interest of ₱5 million, for a total of ₱105 million | Right to receive ₱25 million |

With the setoff, X and Y are still mutual debtors and creditors with respect to the loan and the put option. To settle the loan, we offset their obligations, as follows:

| Parties | Loan and Put Option after setoff |
|---|---|
| X | Right to receive ₱80 million |
| Y | Obligation to pay ₱80 million |

The total amount that X is entitled to receive is computed as follows:

|  | Per Share (₱) | Total (₱) |
|---|---|---|
| Monetary Equivalent of ABC Shares (04 January 2016) | 80 | 80,000,000.00 |
| Less: Strike Price | 105 | 105,000,000.00 |
| Payoff from Put Option | (25) | (25,000,000.00) |
| Add: Loan Receivable |  | 105,000,000.00 |
| Total Amount Entitled (04 January 2016) |  | 80,000,000.00 |

X's entitlement to ₱80 million represents the following: (1) repayment of the principal amount of the loan extended to Y, which is ₱100 million; (2) interest income on the loan, which is ₱5 million; and (3) loss resulting from economic exposure to ABC shares, which is ₱25 million. The amount of ₱80 million is exactly the total value of ABC shares in his portfolio if he purchased 1 million shares.

Y neither gains nor loses any amount, computed as follows:

|  | Per Share (₱) | Total (₱) |
|---|---|---|
| Strike Price | 105 | 105,000,000.00 |
| Less: Monetary Equivalent of ABC Shares (04 January 2016) | 80 | 80,000,000.00 |
| Payoff from Put Option | 25 | 25,000,000.00 |
| Add: New Share Price of ABC Shares (04 January 2016) | 80 | 80,000,000.00 |
| Gross Position | 105 | 105,000,000.00 |
| Less: Loan Repayment |  | 105,000,000.00 |
| Net Position |  | 0 |

Now suppose that the new ABC share price is equal to the strike price. The unrealized gain is ₱5 per share, so that the new share price is ₱105 per share, computed as follows:

|  | Per Share (₱) | Total (₱) |
| --- | --- | --- |
| Share Price (04 January 2015) | 100 | 100,000,000.00 |
| Less: Unrealized Gain | 5 | 5,000,000.00 |
| Share Price (04 January 2016) | 105 | 105,000,000.00 |

Since the new share price of ₱105 per share is equal to the strike price of ₱105 per share, neither party exercises his respective option, letting the option to expire. X does not have an obligation to pay the strike price and Y does not have the obligation to deliver the monetary equivalent of the new share price.

Accordingly, X has zero payoff from either option. Nevertheless, he receives the full payment of the loan extended to Y at maturity date. The total entitlement of X is computed as follows:

|  | Per Share (₱) | Total (₱) |
| --- | --- | --- |
| Monetary Equivalent of ABC Shares (04 January 2016) | 105 | 105,000,000.00 |
| Less: Strike Price | 105 | 105,000,000.00 |
| Payoff from Either Option | 0 | 0.00 |
| Add: Loan Receivable |  | 105,000,000.00 |
| Total Amount Entitled (04 January 2016) |  | 105,000,000.00 |

**Y still does not gain or lose anything:**

|  | Per Share (₱) | Total (₱) |
| --- | --- | --- |
| Strike Price | 105 | 105,000,000.00 |
| Less: Monetary Equivalent of ABC Shares (04 January 2016) | 105 | 105,000,000.00 |
| Payoff from Either Option | 0 | 0.00 |
| Add: New Share Price (04 January 2016) | 105 | 105,000,000.00 |
| Gross Position | 105 | 105,000,000.00 |
| Less: Loan Repayment | 105 | 105,000,000.00 |
| Net Position | 0 | 0.00 |

The point of these illustrations is to establish that, through the combination of the loan and the call and put options, X can replicate stock ownership in ABC shares, even though title over the stock did not transfer to him. On the other hand, Y became indifferent to the change in the share price of ABC shares, as if he ceased to be an owner of the stock.

**Economic Explanation of Constructive Trading**

Without the option contracts, X is simply a creditor to a loan agreement, and Y is a stockholder entitled to residual returns in surplus corporate profits.[115] As creditor and stockholder, they enjoy a set of economic rights provided by default under law,[116] as follows:

|  | **ABC Shares Held by Y** | **Loan Agreement Extended by X** |
|---|---|---|
| **Legal character of security** | Equity instrument | Debt instrument |
| **Risk exposure of original investment** | Exposure to capital gain or capital loss of ABC shares | Security of principal |
| **Income stream** | Variable and non-guaranteed returns | Fixed and guaranteed interest income |
| **Participation rights** | Voting and other control rights | No voting and other control rights |

With the option contracts, some (but not all) of the economic rights pertaining to their respective financial instruments have been exchanged,[117] as follows:

|  | **ABC Shares Held by Y, coupled with Option Contracts** | **Loan Agreement Extended by X, coupled with Option Contracts** |
|---|---|---|
| **Legal character of security** | Equity instrument | Debt instrument |
| **Risk exposure of original investment** | Virtual security of principal | Exposure to capital appreciation and capital loss of ABC shares |
| **Income stream** | Virtual interest income | Exposure to variable and non-guaranteed returns of ABC shares |
| **Participation rights** | Voting and other control rights | No voting and other control rights |

---

115 *Supra* note 9.

116 Pratt, K., *The Debt-Equity Distinction in a Second-best World*, 53 VANDERBILT LAW REVIEW 4, 1055 (2000); Emmerich, A.O., *Hybrid Instruments and the Debt-Equity Distinction in Corporate Taxation*, 52 THE UNIVERSITY OF CHICAGO LAW REVIEW 1, 118 (1985)

117 Hu, H.T. and Black, B., *Debt, Equity and Hybrid Decoupling: Governance and Systemic Risk Implications*, 14 EUROPEAN FINANCIAL MANAGEMENT 4, 663 (2008)

In finance, this scenario can be explained through the put-call parity theorem. The theorem provides that, given a particular time period, the *return* on a share of stock *plus* the *payoff* from an option to sell the share of stock is equivalent to the *return* on a loan agreement *plus* the *payoff* from an option to buy the share of stock. This is expressed as follows:

$$\text{Return on Shares of Stock} + \text{Payoff from Option to Sell}$$
$$= \text{Return on Loan Agreement}$$
$$+ \text{Payoff from Option to Buy}$$

Two features are noteworthy. First, the four terms of the equation speak of "returns" and "payoffs". Second, the left side of the equation involves an equity instrument while the right side involves a debt instrument. Therefore, the theorem describes a situation where holding an equity instrument is economically equivalent to holding a debt instrument, given the proper combination of two option contracts.[118]

For this equation to be true, the following conditions[119] must be satisfied:

1. The strike price of the call option must be equal to the strike price of the put option;

2. The strike price of each option must be equal to the principal of the loan and interest at maturity;

3. The principal of the loan must be equal to the beginning price of the share of stock;

4. The call and put option must have the same underlying asset, which is the share of stock;

5. The call and put options must both be European-style options; and

6. The exercise date of either option must be the same as the maturity date of the loan.

Thus, a party seeking to replicate the economic profile of an equity instrument, *without holding the equity instrument*, can hold a debt instrument, provided these six conditions exist. This is illustrated as follows:

$$\text{Return on Share of Stock}$$
$$= \text{Return on Loan Agreement} + \text{Payoff from Option to Buy}$$
$$- \text{Payoff from Option to Sell}$$

Similarly, a party seeking to replicate the economic profile of a debt instrument, *without holding the debt instrument*, can hold an equity instrument, given the six conditions. This is illustrated as follows:

---

118 *Supra* note 9.
119 *Id.*

$$\begin{aligned}\text{Return on Loan Agreement} &= \text{Return on Share of Stock} + \text{Payoff from Option to Sell} \\ &\quad - \text{Payoff from Option to Buy}\end{aligned}$$

These mathematical relationships show that while the law classifies transactions in clear and categorical terms, it is possible to have two transactions with different legal classifications and different regulatory treatments, but have the same economic characteristics. The more calculating trader or investor will always choose the less onerous classification and treatment as a way to evade regulatory restrictions.

**Conclusion**

Through the proliferation of derivative contracts in the financial system, investors and market players acquire the tools to unbundle economic interest from stock ownership. This has significant ramifications in several areas of regulation, and insider trading is only one area in which derivatives can sidestep the State policy behind these regulations.

In foreign investments law and corporate nationality, for instance, a foreign investor prohibited from purchasing shares of stock in a corporation engaged in a nationalized economic activity, can replicate the economic profile of the shares of stock without acquiring title over the prohibited shares through the same transaction we demonstrated for constructive insider trading. Moreover, in the law on foreign ownership of land, a foreigner can obtain economic exposure in the increased prices of land without owning or buying lands, the ownership of which are reserved to Filipino citizens. In the rules on beneficial ownership in securities regulation, a holder can enjoy the cash flow stream of a security without triggering the acquisition of at least 5% of issued shares of stock in a corporation, and therefore avoiding the beneficial ownership reporting obligation in the Securities Regulation Code. In the capital gains tax system, a taxpayer can cash out or monetize his unrealized gains on a capital asset without triggering a taxable realization event, allowing him to defer capital gains tax liability to another period. In general banking law, a bank that has maximized its equity investment limitation in an allied or non-allied enterprise can still acquire interests that simulate equity investments in said enterprise without acquiring additional shares. In Islamic financing, an Islamic financial institution, which is prohibited by *Shar'ia* from exacting interest income from clients, can simulate interest-like returns on *Shar'ia*-compliant products.

Constructive trading through derivative contracts is a derogation of the State Policy in Section 2 of the Securities Regulation Code, which mandates that "[t]he State shall […] minimize if not totally eliminate insider trading and other fraudulent or manipulative devices and practices which create distortions in the free market." The Supreme Court in *SEC vs. Interport Resources Corp.*[120] provides guidance on how to construe the insider trading law by stating that "'the broad language of the anti-fraud provisions,' which include the provisions on insider trading, should not be 'circumscribed by fine distinctions and rigid classifications.' The ambit of anti-fraud provisions is necessarily broad so as to embrace the infinite variety of deceptive conduct." And it is submitted that constructive trades belong to the "infinite variety of deceptive conduct" that the insider trading law is meant to regulate.

---

120 G.R. No. 135808, October 06, 2008